\documentclass[a4 paper, 12 pt] {article}

\begin{document}
\title{\bf{Massive or Massless Scalar Field and Confinement}}
\author{M. \'{S}lusarczyk $^{a,b)}$ \thanks{mslus@phys.ualberta.ca} \, and A.
Wereszczy\'{n}ski $^{a)}$ \thanks{wereszcz@alphas.if.uj.edu.pl}
       \\
       \\ $^{a)}$ Institute of Physics,  Jagiellonian University,
       \\ Reymonta 4, Krakow, Poland
       \\
       \\  $^{b)}$ Department of Physics, University of Alberta,
       \\ Edmonton, Alberta, Canada T6G 2J1}
\maketitle
\begin{abstract}
Generalization of QCD motivated classical SU(2) Yang--Mills theory
coupled to the scalar field is discussed. The massive scalar
field, corresponding the scalar glueball, provides a confining
potential for static, point--like, external sources. In case of
massless scalar field screening solutions are found. However,
there is a confining sector as well. Both, massive and massless
confining solutions are compared with phenomenological potentials.
The case of non-dynamical permittivity is also discussed.
\end{abstract}
%%%%%%%%%%%%%%%%%%%%%%%%%%%%%%%%%%%%%%%%%%%%%%%%%%%%%%%%%%%%%%%%%%%%%
\section{\bf{The model}}
%%%%%%%%%%%%%%%%%%%%%%%%%%%%%%%%%%%%%%%%%%%%%%%%%%%%%%%%%%%%%%%%%%%%%
Recently, Dick and Fulcher \cite{dick3} have proposed an
interesting model where the lowest glueball, represented by a
massive scalar field $\phi $, has been in a non-minimal way
coupled to the $SU(2)$ gauge fields. They have chosen, in analogy
with the chiral quark models \cite{chiral}, one--glueball coupling:
$\phi F^{a \mu \nu}F^a_{\mu \nu }$. Their model provides
confinement of quarks and gives a quite reasonable interquark
potential $U_{q\bar{q}} \sim r^{1/3}$.
\newline
However, as it was pointed out by Zalewski and Motyka, the most
probable potential, which gives the best fit to the quarkonium
system takes a different form \cite{motyka}, namely
\begin{equation}
U_{MZ}(r)=C_1 \left(\sqrt{r} -\frac{C_2}{r} \right),
\label{potkacper}
\end{equation}
where $C_1\simeq 0.71$ Gev$^{\frac{1}{2}}$ and $C_2 \simeq 0.46$
Gev$^{\frac{3}{2}}$. Due to that the model should be modified. It
can be done by using slightly more complicated, effective coupling. It
has to be underlined that in spite of that modification the scalar
field still represents the lowest scalar glueball.
\newline
In the present paper we focus on the following scalar--gauge action
\begin{equation}
S = \int d^{4}x \left[ -\frac{1}{4} \frac{
(\frac{\phi}{\Lambda})^{8\delta}}{1+(\frac{\phi}{\Lambda})^{8\delta}}
 F^{a}_{\mu
\nu} F^{a \mu \nu} + \frac{1}{2} \partial_{\mu} \phi
\partial^{\mu} \phi  - \frac{m^2}{2} \phi^2 \right].
\label{model}
\end{equation}
Here, $m$ is a mass of the scalar field, $\Lambda $ is a
dimensional and $\delta $ a dimensionless constant. In order to
have the standard Maxwell--like behavior of the fields in the
neighborhood of sources the particular form of the denominator has
been added. It does not inflect the long range behavior of the
fields but provides the asymptotic freedom in the limit of the
strong scalar field \cite{dick2}.
\newline
The field equations corresponding to the action (\ref{model}) take
the form
\begin{equation}
 D_{\mu} \left( \frac{
(\frac{\phi}{\Lambda})^{8\delta}}{1+(\frac{\phi}{\Lambda})^{8\delta}}
F^{a \mu \nu} \right)= j^{a \nu}, \label{f_eq_1}
\end{equation}
and
\begin{equation}
\partial_{\mu} \partial^{\mu} \phi = -2\delta F^{a}_{\mu \nu} F^{a \mu \nu}
\frac{ \phi^{8 \delta - 1}} {\Lambda^{8\delta }\left( 1+\frac{ \phi^{8
\delta}}{\Lambda^{8\delta }}\right)^2} -m^2 \phi , \label{f_eq_2}
\end{equation}
where $j^{a \mu }$ is an external current.
\newline
\newline
In the next section it will be shown that the action (\ref{model})
possesses confining solutions in the massive sector. Moreover,
these solutions describe a wide family of confining potentials
which can be compared with various potentials found in fits to the
phenomenological data \cite{cornel}, \cite{potential},
\cite{martin}.
\newline
Interestingly enough, there is also a confining solution in the
massless sector, where the scalar field cannot be identified with
any glueball. Finally we prove that model with the non-dynamical
scalar field provides confinement of external sources as well.
Problem of glueball states in these models is discussed in the
last section.
%%%%%%%%%%%%%%%%%%%%%%%%%%%%%%%%%%%%%%%%%%%%%%%%%%%%%%%%%%%%%%%%%%%%%
\section{\bf{Confining solutions}}
%%%%%%%%%%%%%%%%%%%%%%%%%%%%%%%%%%%%%%%%%%%%%%%%%%%%%%%%%%%%%%%%%%%%%
%%%%%%%%%%%%%%%%%%%%%%%%%%%%%%%%%%
\subsection{\bf{Massive case }}
%%%%%%%%%%%%%%%%%%%%%%%%%%%%%%%%%%
Let us now discuss the Coulomb problem. We find solutions of the
field equations generated by an external, static, point-like
source:
\begin{equation}
j^{a \mu} = 4\pi q \delta(r) \delta^{a3} \delta ^{\mu 0}.
\label{source}
\end{equation}
The restriction to the Abelian source is not essential. It is
possible to investigate more general, static non-Abelian source
\cite{dick1}: $j^a_{\mu } = 4\pi q \delta (r) C^a \delta^{\mu 0}$.
Here $C^a $ is the expectation value of the $su(N_c)$ generator
for a normalized spinor in color space. However, fields generated
by above sources  have identical dependence on the spatial
coordinates. They differ only by a multiplicative color factor.
Due to that it is sufficient to analyze only Abelian source
(\ref{source}). Then the field equations read
\begin{equation}
\left[ r^2 \frac{
\left(\frac{\phi}{\Lambda}\right)^{8\delta}}{1+\left(\frac{\phi}{\Lambda}\right)^{8\delta}}
E \right]' =4\pi q \delta(r), \label{culomb1}
\end{equation}
\begin{equation}
\nabla^2_r \phi =-4\delta E^2 \frac{ \phi^{8 \delta - 1}}
{\Lambda^{8\delta }\left(1+\frac{ \phi^{8
\delta}}{\Lambda^{8\delta }}\right)^2} + m^2 \phi. \label{culomb2}
\end{equation}
We use the standard definition $E^{ai}=-F^{a0i}$. The electric
field is chosen in the same color direction as the source
$\vec{E}^a=E (r) \delta^{3a} \hat{r}$. One can easily derive the
electric field  from equation (\ref{culomb1}) and express it in
terms of the scalar field:
\begin{equation}
E(r)=\frac{q}{r^2} \left( 1 + \left( \frac{\phi}{\Lambda }
\right)^{ -8\delta } \right). \label{poleE}
\end{equation}
This field can be understood as usual Coulomb field $E(r)
=\frac{\alpha_{eff}}{r^2 }$ in unusual medium. The medium is
characterized by the scalar field which modifies the effective coupling
constant $\alpha_{eff}$:
\begin{equation}
\alpha_{eff}=q \left( 1 + \left( \frac{\phi}{\Lambda } \right)^{
-8\delta } \right). \label{alfa}
\end{equation}
Using relation (\ref{poleE}) we are able to rewrite the equation
(\ref{culomb2}) in the following form:
\begin{equation}
\nabla^2_r \phi =-4\delta \frac{q^2}{\Lambda r^4} \left(
\frac{\phi}{\Lambda } \right)^{ -8\delta -1} +m^2 \phi.
\label{polePhi}
\end{equation}
Unfortunately, this equation is still too complicated to integrate it
analytically. However, on account of the fact that we
are mainly interested in the long range behavior of the fields we
can analyze (\ref{polePhi}) in the asymptotic regime i.e. for $r
\rightarrow \infty $. The asymptotic solution is found to be:
\begin{equation}
\phi \sim \left( \frac{2 \sqrt{\delta } q \Lambda^{4\delta
} }{m} \right)^{\frac{1}{1+4\delta }} \left( \frac{1}{r}
\right)^{\frac{2}{1+4\delta }}. \label{asysol1}
\end{equation}
Then the electric field:
\begin{equation}
E \sim \frac{q}{r^2} + \left( \frac{4 \delta q}{m^2}
\right)^{\frac{-4\delta }{1+4\delta }} \Lambda^{\frac{8\delta
}{1+4\delta }} \left( \frac{1}{r} \right)^{\frac{ 2-8\delta
}{1+4\delta }}. \label{asysol2}
\end{equation}
Finally, the corresponding potential has the form:
\begin{equation}
U \sim  -\frac{q}{r} + \frac{1+4\delta }{12\delta -1} \left(
\frac{4 \delta q}{m^2} \right)^{\frac{-4\delta }{1+4\delta }}
\Lambda^{\frac{8\delta }{1+4\delta }}  \left( \frac{1}{r}
\right)^{\frac{ 1-12\delta }{1+4\delta }}, \label{asysol3}
\end{equation}
for $\delta \neq -\frac{1}{4} $ or $ \frac{1}{12}$. One can
observe that $\delta =\frac{1}{12}$ corresponds to logarithmic
behavior of the electric potential:
\begin{equation}
U \sim -\frac{q}{r} + \left( \frac{q}{3m^2} \right)^{-\frac{1}{4}}
\Lambda^{\frac{1}{2}} \ln  r. \label{asysol4}
\end{equation}
In the framework of such classical models the confinement is
understood as appearance of singularity of the electric potential
(energy density) at the spatial infinity. Energy becomes infinite
due to the long range behavior of fields. This effect takes place
for $\delta \geq \frac{1}{12}$. However, it was shown by Seiler
\cite{seiler} that confining potentials cannot grow faster than
linearly for large $r$. This gives us the upper bound for parameter
$\delta $. Concluding, the model given by the action (\ref{model})
can be used to modelling confinement of the external sources for the
following parameter $\delta $:
\begin{equation}
\delta \in \left[ \frac{1}{12}, \frac{1}{4} \right]
\label{warunek1}
\end{equation}
The standard linear potential is obtained for $\delta
=\frac{1}{4}$. Then we achieve a model discussed previously by
Dick \cite{dick2}. In this model a global, confining solution was
found. As one could expect the asymptotic behavior of the solution
is linear.
%%%%%%%%%%%%%%%%%%%%%%%%%%%%%%%%%%
\subsection{\bf{Massless case}}
%%%%%%%%%%%%%%%%%%%%%%%%%%%%%%%%%%
In this section we are going to focus on the problem of the
confinement in case of the massless scalar field. Apparently, such
a field can be no longer identified with the scalar glueball.
Nonetheless, one can still use it to modify properties of the
vacuum. The scalar field becomes an effective field describing
dynamical permittivity of the medium i.e. vacuum. One can notice
that the model belongs to the so--called color dielectric field
theories.
\newline
Equation (\ref{polePhi}) for the scalar field takes  simpler
form now:
\begin{equation}
\nabla^2_r \phi =-4\delta \frac{q^2}{\Lambda r^4} \left(
\frac{\phi}{\Lambda } \right)^{ -8\delta -1}, \label{mpolePhi}
\end{equation}
which can be integrated analytically. We have found a family of
solutions regular at the spatial infinity and labelled by a
positive parameter $\beta_0$:
\begin{equation}
\phi = A\Lambda \left( \frac{1}{r\Lambda } + \frac{1}{\beta_0}
\right)^{\frac{1}{1+4\delta }}, \label{msol1}
\end{equation}
where the constant $A$ is given by
$$ A=[q(1+4\delta )]^{\frac{1}{1+4\delta }}. $$
The corresponding electric field has the form:
\begin{equation}
E= \frac{q}{r^2} + A^{-8\delta }\frac{q}{r^2} \left(
\frac{1}{r\Lambda } + \frac{1}{\beta_0} \right)^{\frac{-8\delta
}{1+4\delta }}. \label{msol2}
\end{equation}
We see that there exists infinite set of electric solutions,
generated by external point source, which falls as $r^{-2}$ for
large $r$. The behavior of the electric field near a source
strongly depends on the value of the parameter $\delta $. If
$\delta \geq 0$ the singularity at $r=0$ is identical to that
generated by point source in the standard Maxwell electrodynamics.
If $\delta <0$ then the singularity can be arbitrarily large.
\newline
Besides the family of solutions mentioned above, there is a
single field configuration, which solves the equations of motion,
and gives confinement. The solution is:
\begin{equation}
\phi = A\Lambda \left( \frac{1}{r\Lambda }
\right)^{\frac{1}{1+4\delta }}. \label{msol3}
\end{equation}
The electric field is then given by the formula:
\begin{equation}
E=\frac{q}{r^2} + A^{-8\delta }q \Lambda^2 \left(
\frac{1}{r\Lambda } \right)^{\frac{2}{1+4\delta }}. \label{sol3}
\end{equation}
Thus the electric potential for $\delta \neq \frac{1}{4}$ takes
the form:
\begin{equation}
U=-\frac{q}{r} + \frac{4\delta +1}{4\delta -1} A^{-8\delta } q
\Lambda^{\frac{8\delta }{1+4\delta }} \left( \frac{1}{r}
\right)^{\frac{1-4\delta }{1+4\delta }} \label{sol4}
\end{equation}
and for $\delta =\frac{1}{4}$
\begin{equation}
U=-\frac{q}{r} + qA^{-8\delta } \Lambda \ln (\Lambda r).
\label{sol5}
\end{equation}
In fact, we observe confining behavior (in the same sense as for
the massive field) for the parameter $\delta$:
\begin{equation}
\delta \in \left[ \frac{1}{4}, \infty \right). \label{warunek2}
\end{equation}
As it was recently shown \cite{my1}, it is possible to get rid of
the non-confining solutions (\ref{msol1}), (\ref{msol2}) adding
the following potential term
\begin{equation}
V(\phi )=\alpha \phi^4 \left(\frac{\phi}{\Lambda}
\right)^{8\delta}, \label{poten}
\end{equation}
where $\alpha $ is a positive constant. Then only the confining
solution survives. Quite interesting, if we omit the denominator
in the dielectric function in the action (\ref{model}) (that means
we are no longer interested in restoration of the standard $r^{-2}$
behavior of the electric field in  small neighborhood of a
source) then the non-confining solutions describe the so--called
screening phenomena known from quantum Yang--Mills theory
\cite{kiskis}. The field configuration generated by a fixed charge
can have arbitrary small but positive energy \cite{my2}.
%%%%%%%%%%%%%%%%%%%%%%%%%%%%%%%%%%
\subsection{\bf{Non-dynamical permittivity}}
%%%%%%%%%%%%%%%%%%%%%%%%%%%%%%%%%%
Let us now proceed and discuss the third possibility that is
to neglect the kinetic term of the scalar field in
(\ref{model}). In this case $\phi $ is no longer a dynamical
field \cite{thooft}. In the other words it is possible to treat it as
an additional field which, after expressing it in terms of the
gauge fields, can be completely removed from the action. One can easily
observe that in order to deal with a non-trivial theory we are forced to
 add a  potential term for the scalar field. In our calculation the
potential is chosen as in the massless case (\ref{poten}).
\newline
Then the action takes the following form:
\begin{equation}
S=\int d^4x \left[ -\frac{1}{4} \left(\frac{\phi}{\Lambda}
\right)^{8\delta}
 F^{a}_{\mu
\nu} F^{a \mu \nu}-\alpha \phi^4
\left(\frac{\phi}{\Lambda}\right)^{8\delta} \right].
\label{nondyn}
\end{equation}
After variation with respect to the scalar field one gets
\begin{equation}
8\delta F \left(\frac{\phi}{\Lambda} \right)^{8\delta -1}
 - 4(2\delta +1) \alpha \phi^4 \left(\frac{\phi}{\Lambda} \right)^{8\delta
 -1}=0,
\label{neq1}
\end{equation}
where $F=-\frac{1}{4} F^{a \mu \nu }F^a_{\mu \nu }$. As previously
we will investigate only the electric part of the theory. Then one
can trivially solve it and find
\begin{equation}
\phi^4 = \frac{E^2}{a}, \label{eliminacja}
\end{equation}
where for simplicity a new constant $a=(2\delta +1)
\alpha $ was defined. Substituting it into the Gauss law we obtain
\begin{equation}
\left[ r^2 \left( \frac{E^2}{a \Lambda^4} \right)^{2\delta } E
\right]'=4\pi q \delta (r). \label{neq2}
\end{equation}
The solution reads
\begin{equation}
E=a^{\frac{2\delta }{1+4\delta }} \left( \frac{|q|}{\Lambda^2 r^2
} \right)^{\frac{1}{1+4\delta }} \Lambda^2. \label{nonsol1}
\end{equation}
This gives us the electric potential
\begin{equation}
U= a^{\frac{2\delta }{1+4\delta }} |q|^{\frac{1}{1+4\delta }}
\frac{4\delta +1}{4\delta -1} \left( \frac{1}{\Lambda r}
\right)^{\frac{1- 4\delta }{1+4\delta }} \Lambda \label{nonsol2}
\end{equation}
for $\delta \neq \frac{1}{4}$ and
\begin{equation}
U=\sqrt[4]{a q^2} \Lambda \ln (\Lambda r) \label{nonsol3}
\end{equation}
for $\delta =\frac{1}{4}$. Identically as for the massless scalar
field the confinement of the point sources is found for all
parameter $\delta $ equal or larger than $\frac{1}{4}$. From the
confinement point of view these models are equivalent.
\newline
As it was said before the action (\ref{nondyn}) can be expressed
only by gauge fields. After a simple calculation one obtains
the Pagels--Tomboulis lagrangian \cite{Pagels}:
\begin{equation}
L=-\frac{1}{4} F^2 \left[ \left( \frac{F^2}{\Lambda^4} \right)^2
\right]^{ \delta }. \label{pagels}
\end{equation}
Of course, the pertinent equations of motion, in the electric
sector, derived for the Pagels--Tomboulis model are the the same as
(\ref{neq2}) (up to unimportant multiplicative constant).
\newline
Models, where the permittivity depends on the strength tensor $F$
has been very successfully analyzed in the context of the
effective theory for the low energy QCD \cite{adler},
\cite{mendel}. Here no additional, effective scalar field is
needed. Confining solutions emerge due to the modification of the
vacuum by the gauge fields. In fact, in the Pagels--Tomboulis model
a source charge generates infinite energy field configuration
whereas a configuration originating from dipol-like source has finite energy
\cite{Pagels}, \cite{my3}.
%%%%%%%%%%%%%%%%%%%%%%%%%%%%%%%%%%%%%%%%%%%%%%%%%%%%%%%%%%%%%%%%%%%%%
\section{\bf{Conclusions}}
%%%%%%%%%%%%%%%%%%%%%%%%%%%%%%%%%%%%%%%%%%%%%%%%%%%%%%%%%%%%%%%%%%%%%
In the present paper we have analyzed the $SU(2)$ gauge field
coupled to the scalar field. For the massive and massless scalar
field as well as for the non-dynamical scalar field solutions
describing confinement have been found. It is relatively easy to
model confinement of external sources using the scalar
dilaton/glueball field. Our models seem to be very similar from
the confinement point of view. Of course, there are also
some differences between them.
\newline
Firstly, the occurrence of confinement strongly depends on the
value of the parameter $\delta $. The massive and massless models
require different values of $\delta $. For example the best
phenomenological quark--antiquark potential, found by Zalewski and
Motyka in the fits to the spectrum of quarkonia, is obtained for
$\delta =\frac{3}{20}$ for the massive and for $\delta
=\frac{3}{4}$ for the  massless case respectively. In fact, then
$U \sim \sqrt{R}$, where $R$ is the distance between sources.
Another known interquark potential $U \sim R^{0.1}$, used by
Martin \cite{martin}, is realized for $\delta =\frac{11}{116}$
(massive field) or for $\delta = \frac{11}{36}$ (massless field).
The difference between these models is even more apparently seen
in case of the linear confinement. The massive model describe it
for $\delta =\frac{1}{4}$. It is unlikely for the massless model,
where the linear potential is realized only in the limit $\delta
\rightarrow \infty $, which cannot be implemented on the action
level. The linear confinement can be only approximated with
arbitrary accuracy  by taking sufficiently large $\delta $.
\newline
To conclude, in spite of the differences mentioned above all the
models considered here can serve  very well to model
confinement of quarks. Nevertheless these models differ
essentially. One of the most profound differences is visible if we
look at the way how glueball states appear in the models.
\newline
\newline
If we would like to treat these models as candidates for description of
low energy QCD, problem of glueballs must be taken into consideration.
It is obvious that a
good effective model is expected  not only to provide confinement of
quarks (and give a reasonable quark--antiquark potential) but is also
expected to have solutions which could be interpreted as glueballs.
In the massive case the situation is clear. The scalar field
represents the scalar glueball $0^{++}$ which has mass $m$. The
dielectric function determines coupling between the glueball and
gluons. For instance, $\delta =\frac{1}{8}$ gives one glueball
coupling. For different values of $\delta $ we have  more complicated,
effective coupling. Unfortunately, this model is unable to
describe other glueball states. In order to do it one has to
enlarge the number of fields. For example the pseudo-scalar
glueball $0^{-+} $, represented by a new field $\psi $, requires
axion-like coupling  \cite{cornwall1}.
\newline
Situation changes drastically if we consider remaining models
(with the dynamical or non-dynamical scalar field). Obviously, the
interpretation that the scalar field corresponds to the glueball
is no longer correct. Because of that, glueballs should appear in
a different,  non-trivial way. It is believed that in the framework
of these models glueballs could be described as (topological)
solitons i.e. static, localized and finite energy solutions of the
sourcesless field equations. This is a very attractive way of studying
glueballs. Different glueballs would be given by
different solitons characterized by a topological index (cf. e.g.
 the Faddeev--Niemi model \cite{niemi}).
\newline
As a result, existence of topologically non-trivial solutions in
the framework of scalar--gauge models is crucial from the QCD
point of view. It would let us justify which model can be really
relevant as the effective model for the low energy quantum
chromodynamics. We plan to address this problem  in our proceeding
paper.

\end{document}